\documentclass[a4paper,conference]{IEEEtran}

\usepackage{graphicx} 
\usepackage{amsmath} 
\usepackage{algorithm}
\usepackage{algpseudocode}
\DeclareMathOperator*{\argmin}{argmin}
\usepackage{pgfplots}
\usepackage{pgfplotstable}
\pgfplotsset{compat=1.18}
\pgfplotscreateplotcyclelist{default}{
    {blue, mark=*},
    {red, mark=square*},
    {brown!60!black, mark=triangle*},
    {black, mark=diamond*},
    {blue!50!red, mark=pentagon*}
}
\pgfplotsset{cycle list name=default}
\usepackage{subcaption}
\usepackage{caption}
\usepackage{comment}

\title{An Online Fragmentation-Aware GPU Scheduler for Multi-Tenant MIG-based Clouds}
\author{Marco Zambianco, Lorenzo Fasol, Roberto Doriguzzi-Corin}
\author{
    \IEEEauthorblockN{Marco~Zambianco\IEEEauthorrefmark{1}, Lorenzo Fasol\IEEEauthorrefmark{1}, Roberto Doriguzzi-Corin\IEEEauthorrefmark{1}}
        \IEEEauthorblockA{\IEEEauthorrefmark{1}Fondazione Bruno Kessler (FBK), Trento, Italy
\\Email: mzambianco, lfasol, rdoriguzzi@fbk.eu}       
}
\date{October 2025}

\pgfplotstableread[col sep=space]{figures/active_gpus.txt}\activegpus
\pgfplotstableread[col sep=space]{figures/acceptance_rate.txt}\acceptancerate
\pgfplotstableread[col sep=space]{figures/tot_scheduled_wl.txt}\scheduledwl
\pgfplotstableread[col sep=space]{figures/res_utilization.txt}\resutil

\begin{document}

\maketitle

\begin{abstract}
The explosive growth of AI applications has created unprecedented demand for GPU resources. Cloud providers meet this demand through GPU-as-a-Service platforms that offer rentable GPU resources for running AI workloads. In this context, the sharing of GPU resources between different tenants is essential to maximize the number of scheduled workloads. Among the various GPU sharing technologies, NVIDIA’s Multi-Instance GPU (MIG) stands out by partitioning GPUs at hardware level into isolated slices with dedicated compute and memory, ensuring strong tenant isolation, preventing resource contention, and enhancing security. Despite these advantages, MIG’s fixed partitioning introduces scheduling rigidity, leading to severe GPU fragmentation in multi-tenant environments, where workloads are continuously deployed and terminated. Fragmentation leaves GPUs underutilized, limiting the number of workloads that can be accommodated.
To overcome this challenge, we propose a novel scheduling framework for MIG-based clouds that maximizes workload acceptance while mitigating fragmentation in an online, workload-agnostic setting. We introduce a fragmentation metric to quantify resource inefficiency and guide allocation decisions. Building on this metric, our greedy scheduling algorithm selects GPUs and MIG slices that minimize fragmentation growth for each incoming workload.
We evaluate our approach against multiple baseline strategies under diverse workload distributions. Results demonstrate that our method consistently achieves higher workload acceptance rates, leading to an average 10\% increase in the number of scheduled workloads in heavy load conditions, while using approximately the same number of GPUs as the benchmark methods.
\end{abstract}

\begin{IEEEkeywords}
Multi-instance GPU, GPU fragmentation, resource allocation
\end{IEEEkeywords}
\section{Introduction}

The rapid proliferation of AI applications has created an unprecedented demand for GPUs capable of provisioning a wide range of services (i.e., chatbots, computer vision tasks, recommendation systems, etc.) \cite{miao2025towards}. To streamline the deployment process of these applications and reduce capital expenditure sustained by service providers to buy the necessary GPU hardware, cloud providers offers GPU-as-services platforms which enable the renting of GPU resources based on a pay-per-use business model. In this scenario, GPU sharing techniques are often used to multiplex applications of various tenants within the same GPU hardware, improving the resource utilization efficiency and thus boosting renting revenues for the cloud provider \cite{choi2022serving}. In contrast to traditional compute resources (e.g., CPU, memory, and storage), which can be readily shared among co-located workloads, GPUs necessitate specialized techniques such as time slicing \cite{garg2018share}, multi-process service (MPS) \cite{dhakal2020gslice}, and GPU virtualization (vGPU) to support concurrent sharing.  Among these, NVIDIA’s proprietary Multi-Instance GPU (MIG) feature, which is increasingly supported by new generation graphic cards, distinguishes itself by allowing a spatial partition of a GPU in multiple slices with dedicated compute and memory resources \cite{zhao2021survey}. In multi-tenants environment, this sharing strategy is an appealing option compared to logical sharing techniques performed at software-level as it makes it possible to fully isolate tenants allocated on the same GPU. Specifically, hardware-level isolation prevents resource contention between workloads \cite{li2022characterizing} and improves security by shielding them from side-channel and denial-of-service attacks initiated  by co-located malicious applications \cite{pavlidakis2024guardian}. However, these benefits come at the cost of scheduling flexibility since partition options are restricted to predetermined profiles which identify specific compute and memory slices of the GPU.  As a result, finding strategies to effectively allocate workloads according to these constraints becomes fundamental to maximize the number of accommodated applications. 

\begin{figure}[t]
    \centering
    \begin{subfigure}[b]{1\linewidth}
        \centering
        \includegraphics[width=1\linewidth, trim={0cm 1.5cm 0cm 0.5cm}, clip]{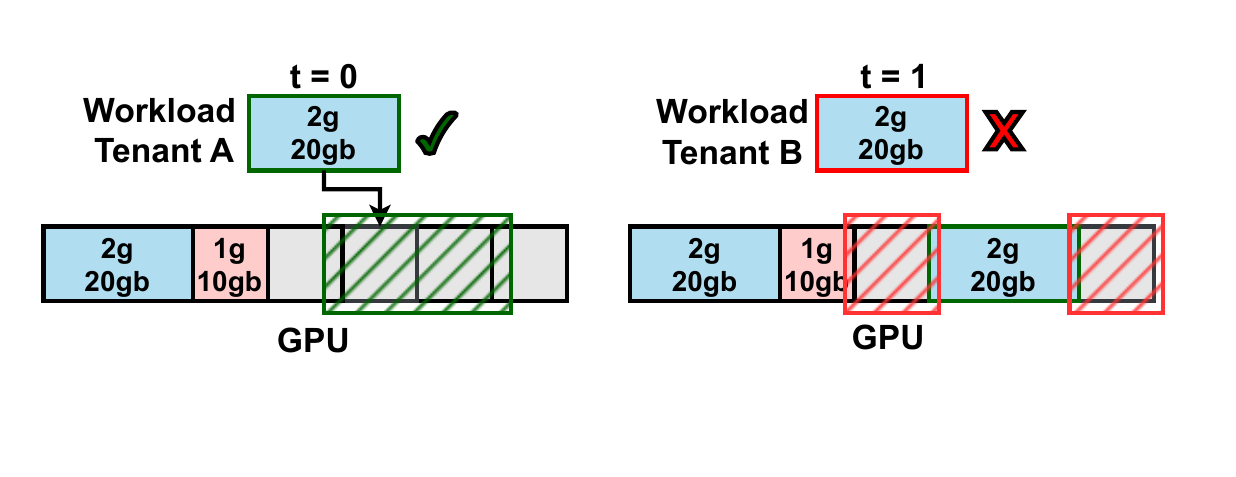}
        \caption{GPU fragmentation caused by workload arrival. }
        \label{subfig:arrival_frag_example}
    \end{subfigure}
    
    \begin{subfigure}[b]{1\linewidth}
        \centering
        \includegraphics[width=1\linewidth]{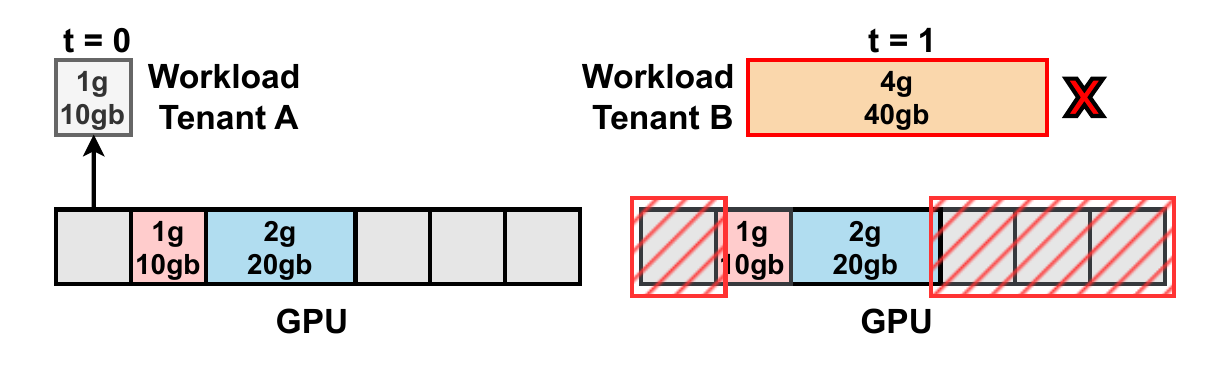}
        \caption{GPU fragmentation caused by workload termination.}
        \label{subfig:termination_frag_example}
    \end{subfigure}

    \caption{Examples of GPU fragmentation caused by workload arrival and termination. Fragmentation is represented by red hatched rectangles.}
    \label{fig:arrival_termination_frag_examples}
\end{figure}

In this context, the research activity has proposed different solutions to improve the resource utilization of MIG-based deployments such as the computation of the optimal partition size that minimize resource waste and the adoption of logical slicing to access unused resources within each MIG partition \cite{wei2024optimizing,zhu2024optimizing,qi2024elasticbatch}. 
These techniques assume a single-tenant setup, where one service provider controls both the life cycle of its AI services and the resources of underlying GPU infrastructure. In this setting, the provider can combine several of the earlier solutions to work around MIG limitations. However, in a multi-tenant environment, the cloud provider that schedules applications on shared GPUs usually has little insight into each workload’s behavior or lifespan. Consequently, such solutions cannot be safely applied without risking disruptions to the availability of running applications.
Moreover, the constant deployment and termination of workloads, combined with MIG’s rigid scheduling rules, leads to heavy fragmentation of GPU resources. The effect of fragmentation reduces the number of schedulable workloads on a GPU because the available MIG slices do not comply with the partition constraints and cannot be utilized. To better illustrate this aspect, we provide a pictorial representation in Fig. \ref{fig:arrival_termination_frag_examples} of the fragmentation dynamic when new workloads are scheduled (Fig. \ref{subfig:arrival_frag_example}) and when they are terminated (Fig. \ref{subfig:termination_frag_example}).
Building on these observations, we aim to adapt MIG-based deployments to multi-tenant scenarios while addressing the issue of GPU resource fragmentation. To this end, we approach fragmentation mitigation from a resource allocation perspective and design an online scheduling algorithm for MIG profiles that effectively reduces GPU fragmentation, without prior knowledge about workload resource distribution. As a result, by improving the GPU utilization, a greater number of workloads can be successfully scheduled, thereby increasing the potential renting revenue for the cloud provider. In detail, the main contribution of this work are as follows:
\begin{itemize}
    \item We design a metric to analytically measure the fragmentation severity experienced by GPUs shared using MIG. This metric makes it possible to compare fragmentation levels across GPUs based on their currently allocated MIG profiles and can be used to support more informed decisions about future workload placements.
    \item We design a scheduling algorithm based on a greedy approach that leverages the proposed fragmentation metric to maximize the workload acceptance rate. For each workload to schedule, which requests GPU resources in the form of a MIG profile, the algorithm evaluates the current fragmentation severity of the available GPUs and select the GPU and related MIG slices that minimize the fragmentation growth.
    \item We evaluate the performance of the proposed scheduling solution with respect to several benchmark schemes using different strategies to select GPUs and related MIG slices. The results show that our approach  outperforms the baselines in all metrics and provides consistent performance regardless of the workload MIG profile distribution.
\end{itemize}

The remainder of the paper is organized as follows. In Section II, we discuss the related work. Section III provides some background information on the main concepts of MIG.  Section IV describes the considered system model. Section V presents the fragmentation-aware algorithm for MIG-based deployments. In Section VI, we discuss the obtained results. Finally, we draw the conclusion in Section VII.
\section{Related Work}



The problem of scheduling of AI workload on multi-tenant GPU clusters has been broadly investigated in the recent years. 
An overview of the main concepts and related challenges can be found in \cite{fu2025survey} and \cite{ye2024deep}. 
                                       
                                       
The authors of \cite{li2022miso} tackles MIG slice under-utilization using MPS to estimate appropriate resource needs and select optimal MIG partitions to improve AI service performance. Similarly, \cite{zhang2024miger} and \cite{lee2024parvagpu} integrate MIG with MPS to boost utilization in underutilized slices while controlling  resource contention caused by logical slicing. In contrast, \cite{villarrubia2025solving} employs a reinforcement learning agent to minimize the makespan of AI workloads on MIG by modeling workload dynamics. Different from these studies, we focus on MIG-based workload allocation from a cloud provider’s perspective, designing a fragmentation-aware scheduling algorithm that improves GPU cluster utilization without relying on external information about workload type and service performance.

The authors of \cite{weng2023beware} analyze the problem of fragmentation in GPU clusters and propose a fragmentation metric for GPU nodes in order to design a fragmentation-aware allocation algorithm. Similarly, the authors of \cite{lao2025cafgd} and \cite{lettich2025power} extend this fragmentation-aware approach to online settings where workload arrival is distributed in time and by including the minimization of GPU power consumption, respectively. Although these solutions address the problem of GPU fragmentation, the proposed fragmentation metrics assume a MPS-based GPU sharing, leaving the MIG-based sharing uncovered. We instead propose a fragmentation metric tailored to MIG which quantifies the fragmentation severity by evaluating the amount of unschedulable MIG profiles.  
                                            
Finally, \cite{siavashi2025multi, turkkan2024optimal} tackle the problem of resource efficiency in MIG-based deployments  designing rescheduling and admission control schemes, respectively, to reduce GPU fragmentation and increase the amount of successfully allocated workloads. However, \cite{siavashi2025multi} uses a static, periodic rescheduling approach, whereas \cite{turkkan2024optimal} relies on prior knowledge of MIG profile distributions to guide its admission control policy. In contrast, our algorithm serves workload requests online, continuously mitigating fragmentation without any a-priori assumptions about the statistic of the requested MIG profiles.

\section{Multi-instance GPU (MIG) background}

MIG is a feature introduced with NVIDIA’s Ampere architecture, and now available on the latest GPUs, that allows multiple jobs to securely share a single GPU without competing for the same resources \cite{nvidia-mig}. By partitioning a physical GPU into up to seven fully isolated instances, each with its own compute and memory resources, MIG allows for efficient multi-tenant usage.
It is particularly effective for AI inference workloads, multi-tenant cloud environments, and scenarios where individual jobs do not fully utilize GPU capacity.

The partitioning mechanism is based on fundamental building blocks called slices. GPU memory is divided into eight memory slices, each representing 1/8 of the total memory capacity. Similarly, streaming multiprocessors (SMs) are divided into seven SM slices, each containing roughly 1/7 of the total compute units. Typically, a GPU slice is formed by pairing one memory slice with one SM slice, except for the last GPU slice, which combines one SM slice with two memory slices. 

\begin{figure}[t]
    \centering
    \includegraphics[width=\linewidth]{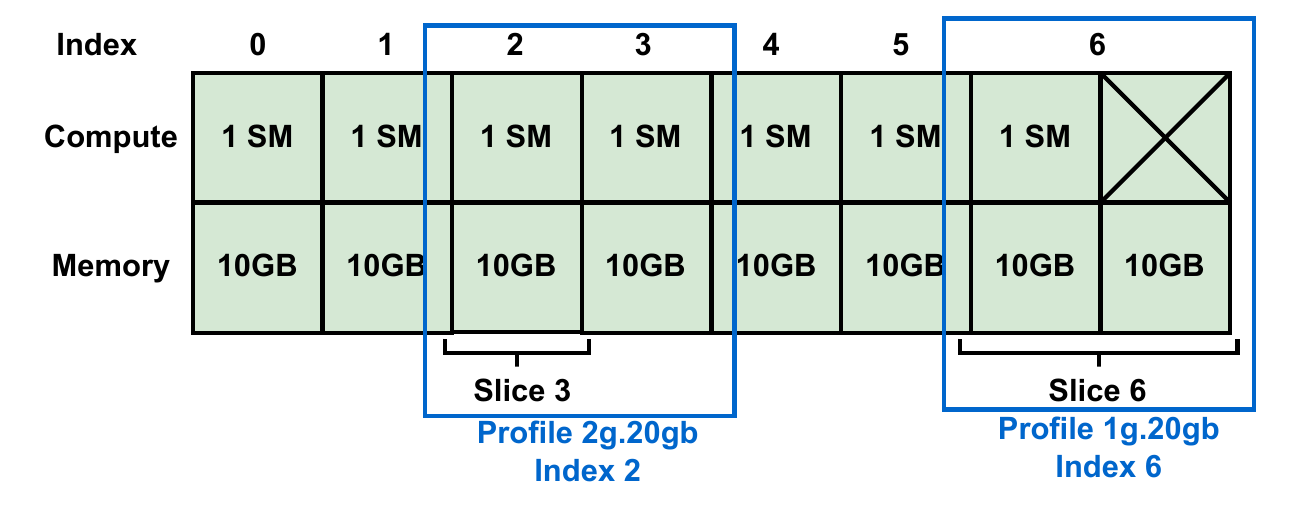}
    \caption{A100 MIG-based representation of slices and profiles.}
    \label{fig:gpu_representation}
\end{figure}


To standardize and simplify the GPU partitioning, NVIDIA defines several MIG profiles, which describe valid combinations of compute and memory resources that can be exclusive accessed, ensuring complete hardware-level isolation through separate memory pathways. Each MIG profile specifies the size of the compute  and memory slices, using the naming convention \textless{}g\textgreater{}g.\textless{}mem\textgreater{}gb.
For example, as depicted in Fig.  \ref{fig:gpu_representation}, on an \textit{A100-80GB} GPU, a 2g.20gb instance profile allocates two SM slices and two 10GB slices, while a 1g.20gb instance profile uses one SM slices and two 10GB of memory. Note that this procedure can be extended to any GPU model supporting MIG.
Placement indexes determine the physical location on the GPU where each profile is mapped on the hardware, ensuring that the assigned slices are contiguous.  We resume all possible feasible profile configurations for the A100 model in Tab. \ref{tab1:A100_slice}. This contiguity requirement implies that not all combinations of instance profiles can coexist simultaneously, as available slices must be assigned in contiguous blocks without gaps. The adoption of placement indexes introduces additional constraints on configuration flexibility during the creation and removal of dynamic instances. Consequently, this leads to an optimization problem focused on maximizing resource utilization and minimizing fragmentation, in order to reduce inefficiency and resource wastage.
\section{System model}

\begin{table}[t] 
\centering
\caption{MIG specifications for A100 GPU}
\begin{tabular}{| c | c | c | c | c | c | c | c |}
\hline                                              
\textbf{Profile} & \textbf{Slice} & \textbf{No. Instances} & \textbf{Index} \\                  
\hline
7g.80gb & 7 & 1 & \{0\} \\
\hline
4g.40gb & 4 & 1 & \{0\} \\
\hline
3g.40gb & 4 & 2 & \{0, 4\}\\
\hline
2g.20gb & 2 & 3 & \{0, 2, 4\}  \\
\hline
1g.20gb & 2 & 4 & \{0, 2, 4, 6\}  \\
\hline
1g.10gb & 1 & 7 & \{0, 1, 2, 3, 4, 5, 6\} \\
\hline
\end{tabular}
\label{tab1:A100_slice}
\end{table}

We consider a homogeneous GPU cluster composed of the same hardware model (e.g. A100, H100, etc) that support MIG. The infrastructure is owned and managed by  a cloud provider which offers a GPU-as-a-Service platform that allows external users to rent GPUs resources based on a pay-per-use billing model. We assume that MIG is employed as GPU sharing technology so each  tenant can request fractional GPU resources tailored to run his AI applications in an isolated multi-tenant environment. Formally, we indicate the set of GPUs in the cluster as $M$, where each GPU $m$ is composed by $S_m$ slices identified by the set of indexes $I = \{0, ..., S_m-1\}$. The amount of supported MIG profiles is indicated by the set $P = \{\text{1g.10gb}, ..., \text{4g.40gb}, \text{7g.80gb}\}$ and the related set of feasible indexes as $I_p \subseteq I, p \in P$. Similarly, we define the set of workloads to schedule as $W$, where each workload $w$ requests a MIG profile $p \in P$ which corresponds to a specific amount of compute and memory GPU resources $r_w(p) = \langle r^{comp}, r^{mem}\rangle, \forall w \in W$.  Workload arrivals are distributed in time and are served by the infrastructure scheduler in an online fashion following a First-In First-Out queue policy. We indicate the amount of allocated slices in each GPU using the indicator variable $x_{m,i}, m \in M,i \in I$, which assumes value 1 if slice $i$ of GPU $m$ is allocated, 0 otherwise. To counteract the natural increase of resource fragmentation caused by the continuous arrival and termination of workloads which is further exacerbated by MIG, the scheduling logic is designed to individually accommodate each workload request in compliance with MIG constraints while mitigating the resource fragmentation. The benefit of this strategy allows to increase the number of accepted workload requests, thereby generating higher rental revenues for the cloud provider. In this context, we design a fragmentation-aware scheduling algorithm based on the following assumptions: 
\begin{itemize}
\item \textbf{Unknown workload statistics}: the cloud provider is agnostic of both the statistic of the requested MIG profiles and lifespan duration. Therefore, the designed scheduler should provide reliable scheduling decisions regardless of the MIG profile requests distribution, hence it works in an online setting by adapting the scheduling decision to the current state of the GPU cluster.
\item \textbf{Dynamic GPU partition reconfiguration}: MIG partitions are automatically created at runtime based on workload requirements by dynamically aggregating MIG slices. This capability, introduced in recent MIG driver versions and leveraged by real-world GPU schedulers\footnote{https://www.theriseunion.com/en/blog/HAMi-2-5-0.html}, enhances resource flexibility as it eliminates the need to pre-configure partitions on the available GPUs.  
\item \textbf{No workload rescheduling}:  although beneficial for de-fragmenting resources, re-scheduling workloads across GPUs may cause service disruption during the migration process, affecting tenants' applications performance. Therefore, we assume that the scheduler exclusively allocates workloads based on the current  availability of MIG slices without the possibility to modify the allocation in the future (we are going to consider rescheduling in a future work to augment the proposed scheduling logic).

\end{itemize}

Our approach adopts a heuristic scheduling strategy, as opposed to an exact optimal formulation, to ensure practical applicability and scalability. This choice is motivated by the online nature of the scheduling problem, which would otherwise require prior knowledge of future workload arrivals and lifespans to globally optimize the workload allocations. Furthermore, the maximization of the number of allocated workloads can be viewed as a variant of the bin packing problem where items (i.e., MIG profiles) are subject to position constraints within bins (i.e., GPUs), as demonstrated in similar MIG scheduling studies \cite{siavashi2025multi}\cite{turkkan2024optimal}. As a result, the combinatorial complexity  makes the computation of optimal solutions impractical for real-world deployments.


\begin{figure*}[t]
    \centering
    \begin{subfigure}[b]{0.49\linewidth}
        \centering
        \includegraphics[width=\linewidth, trim={0cm 0.5cm 0cm 0.5cm}, clip]{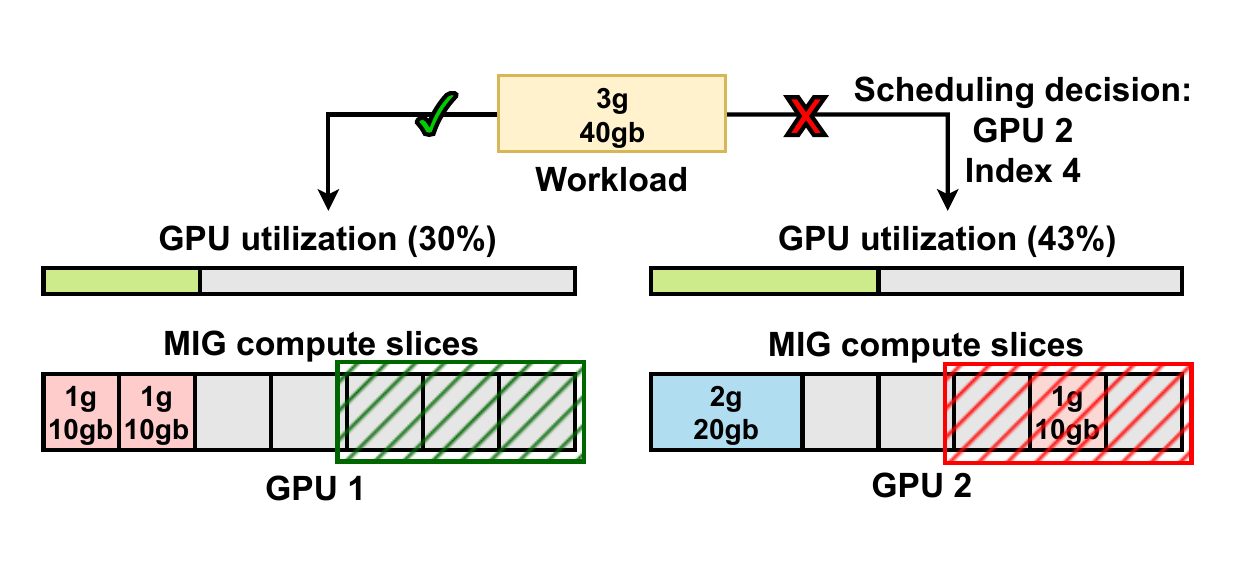}
        \caption{Example of workload rejection using best-fit scheduling }
        \label{subfig:best_fit_rejection}
    \end{subfigure}
    \hfill
    \begin{subfigure}[b]{0.49\linewidth}
        \centering
        \includegraphics[width=\linewidth, trim={0cm 0.5cm 0cm 0.5cm}, clip]{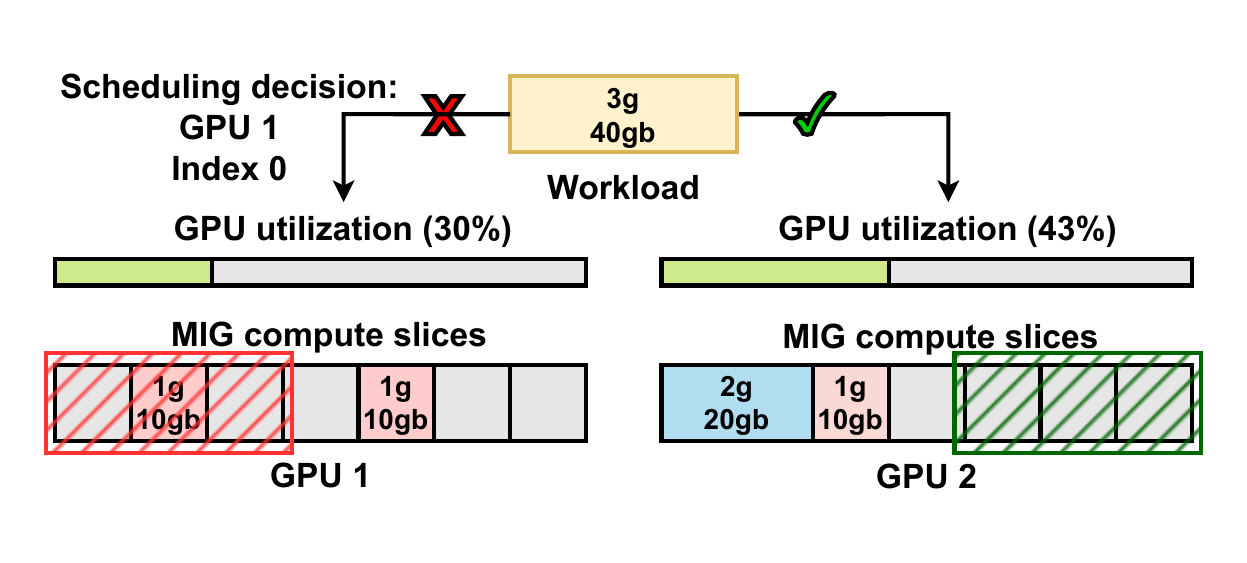}
        \caption{Example of workload rejection using load-balance scheduling }
        \label{subfig:load_bal_rejection}
    \end{subfigure}

    \caption{Examples of workload rejections using different scheduling strategies. Although enough resource are available in both scenarios, the workload cannot be scheduled due to resource fragmentation reducing the amount of allocable MIG slices.}
    \label{fig:rejection_examples}
\end{figure*}

%

\section{Fragmentation-aware scheduling for MIG}
In this section we describe the proposed scheduling algorithm mitigating GPU fragmentation when using MIG. First we define a fragmentation measure for MIG deployments which quantifies the fragmentation severity on GPUs. Then, we leverage this formulation to design a fragmentation-aware allocation scheme of MIG profiles.

\subsection{The impact of MIG fragmentation on scheduling decisions}
Fragmented GPUs can severely alter the scheduling performance by causing unexpected workload rejection. Therefore, the definition of a fragmentation metric tailored to MIG covers an essential role to drive the design of an effective scheduling algorithm overcoming this issue. To highlight the significance of this requirement, Fig. \ref{fig:rejection_examples} presents two examples demonstrating how fragmentation-agnostic scheduling strategies fail to allocate workloads on MIG-based clusters. Specifically, the considered scheduling schemes employ opposite criteria for GPU selection, illustrating this behavior under different scheduling scenarios. One scheme is based on a best-fit logic which selects the GPU minimizing the waste of MIG slices (Fig. \ref{subfig:best_fit_rejection}), the other employs a load-balancing approach and selects the GPU with the least amount of allocated MIG slices (Fig. \ref{subfig:load_bal_rejection}). Both schemes select the MIG slices indexes by searching the first available index which can accommodate the profile. We observe that, although both schemes can identify a GPU with sufficient resources for the workload, the allocation still fails because the available indexes on the selected GPU are incompatible with the requested MIG profile. This problem arises from fragmentation of MIG slices, which reduces the number of effectively schedulable MIG profiles on that GPU.

\subsection{Characterizing fragmentation in MIG}
Resource fragmentation typically refers to the inefficient use of computational resources caused by their non-contiguous allocation. While contiguity is necessary, it alone is not sufficient to determine whether a GPU is fragmented when using MIG, due to various index constraints that limit scheduling options. Consequently, we need to refine the definition of fragmentation to account for these constraints, enabling the creation of a reliable metric that can quantitatively guide scheduling decisions. To this end, we address the following two key questions:
\begin{enumerate}
\item Under what conditions can a GPU employing MIG-based sharing be considered fragmented?
\item Once fragmentation occurs, how can we quantify its severity?
\end{enumerate}

To answer question 1, we argue that the definition of fragmented GPU using MIG is a relative concept that in fact depends on the combination of the MIG profile that is being scheduled and the available indexes. Referring to Fig. \ref{subfig:best_fit_rejection}, workloads requesting the 1g.10gb and 2g.20gb profiles can be successfully scheduled on GPU 2, as sufficient resources and feasible indexes are available, even though some allocated slices are not contiguous. In contrast, the 3g.40gb and 4g.40gb profiles are rejected despite an adequate number of available slices due to feasible indexes being already assigned to other profiles. (see Table \ref{tab1:A100_slice}).
Based on this observation, let $\Delta S_m = S_m - \sum_{i \in I}x_{m,i}$ be the amount of unused slices on GPU $m$:

\textbf{Definition.} \textit{GPU $m \in M$ is fragmented with respect to MIG profile $p$ requested by workload $w$ if $r_w(p) \leq \Delta S_m $ and $\sum_{i \in \{\bar{i}, \bar{i}+1, ..., \bar{i}+ r_w(p) - 1\}} x_{m,i} > 0$, $ \forall  \bar{i} \in I_p$.}

In other words, a GPU is fragmented with respect to profile $p$ if the latter cannot be scheduled although a sufficient amount of slice is available. This definition of fragmentation makes it possible to contextualize its behavior with respect to the scheduling of workloads and can be used to further elaborate a fragmentation measure by answering to question 2. 
As a matter of fact, the proposed definition of fragmentation is too coarse and does not allow to discriminate between different levels of fragmentation severity since a single misplaced MIG profile on an empty GPU would make the GPU fragmented. For example, scheduling profile 1g.10gb on MIG slice at index 1 prevents the allocation of MIG profile 4g.40gb.
To overcome this limitation, we augment the notion of GPU fragmentation by introducing the concept of fragmentation score, referred to as $F(m), m \in M$, which allows to differentiate various level of fragmentation by assessing the scheduling feasibility of the supported MIG profiles. We report pseudo-code to compute the fragmentation score in Algorithm \ref{alg:frag_score}. In detail, given a GPU $m$, we assume that a hypothetical workload $w$ requesting profile $p$ is scheduled. Then, we evaluate the feasible allocation configurations of profile $p$ assessing whether the hypothetical allocation of profile $p$ at index $i \in I_p$ is feasible  (lines \ref{alg:frag_score:frag_check_begin}-\ref{alg:frag_score:frag_check_end}). For each unfeasible allocation, we increment the fragmentation score $F(m)$ by the number of MIG memory slices composing the profile $p$ (line \ref{alg:frag_score:update_frag_score}). This procedure is repeated for each supported profile $p \in P_m$.
\begin{algorithm}[t]
\caption{Fragmentation score computation}\label{alg:frag_score}
\begin{algorithmic}[1]
    \State \textbf{Input:} GPU  $m \in M$
    \State \textbf{Output:} $F(m)$ 
    \For{\textbf{each} profile $p$ in $P_m$}
        \State Assume that workload $w$ requests MIG profile $p$
            \If{$r_w(p) \leq \Delta S_m$}
                \For{\textbf{each} $\bar{i} \in I_p$} \label{alg:frag_score:frag_check_begin} 
                    \If{$ \sum_{i \in \{\bar{i}, \bar{i}+1, ..., \bar{i}+ r_w(p) - 1\}} x_{m,i} > 0$}\label{alg:frag_score:frag_check_end} 
                    \State $F(m) \leftarrow F(m) + r^{mem}$ 
                    \label{alg:frag_score:update_frag_score}
                    \EndIf
                \EndFor 
            \EndIf
    \EndFor
\end{algorithmic}
\end{algorithm}
Intuitively, the fragmentation score associated to a GPU depends on the total number of  unfeasible MIG profile configurations weighted by the number of MIG memory slices composing the profile. Note that we consider memory slices in order to account for the misalignment between compute and memory slices, as the latter are more likely to be fragmented when allocating profiles requiring unequal memory and compute slices such as 1g.20gb and 3g.40gb. The higher the value, the lower is the amount of allocable MIG slices. As a result, it is possible to compare the fragmentation severity produced by different MIG allocations. For example, the fragmentation score of GPU 2 in Fig. \ref{subfig:best_fit_rejection}, would be computed as follows. For each of profile, we compute the number of unfeasible allocations: profile 1g.20gb requires 2 MIG slices (see Table \ref{tab1:A100_slice}) and cannot be allocated on index 4 since the second memory slice is allocated to profile 1g.10gb. As a result, GPU 2 has a partial fragmentation score of 2 (1 unfeasible allocation for a profile composed of 2 MIG memory slices). By repeating this computation for all remaining profiles \{2g.20gb, 3.40gb, 4.40gb\}, the resulting fragmentation score for GPU 2 is $F(2) = 2 + 2 + 8 + 4 = 16$. Following the same procedure, the fragmentation score for GPU 1 is $F(1) = 8$. Therefore, by comparing the fragmentation scores of the two GPUs, we conclude that GPU 2 is more fragmented than GPU 1. This information allows  to analytically compare MIG allocations across GPUs and links resource utilization with the scheduling policy. 



\subsection{Minimum Fragmentation Increment (MFI) algorithm}
We leverage the definition of fragmentation score to design a fragmentation-aware scheduling algorithm that maximizes the number of successfully scheduled workloads (in other words, the number of scheduled MIG profiles that comply with MIG allocation constraints). In an online scenario, where workloads are distributed over time and scheduling decisions depend on the current allocation of GPU resources, fragmentation scores can vary considerably across scheduling intervals. This variation is primarily caused by the limited placement flexibility of certain MIG profiles (e.g., 4g.40gb and 3g.40gb) and by the termination of deployed workloads, which may release GPU resources consisting of non-contiguous MIG slices.
Following this observation, inspired by the allocation strategy proposed by authors of \cite{weng2023beware}, we schedule workloads in order to counteract the fragmentation score growth that is bound to naturally increase over time. Intuitively, this strategy has the effect of reducing the fragmentation and thus increasing the workload scheduling success rate. We implement this idea by designing a scheduling algorithm, named as Minimum Fragmentation Increment (MFI), which adopts a greedy logic to select the GPU and related MIG slices minimizing the increment of fragmentation score. Operationally, the algorithm performs dry-run allocations on each GPU to estimate the variation in fragmentation score performance prior to committing the final scheduling decision. We report the algorithm steps in Algorithm \ref{alg:mfi}. The algorithm performs the following steps for each workload to schedule $w$. We initialize the fragmentation score variation, $\Delta F(m)$, for each GPU $m$ which is going to be used to assess the impact on fragmentation caused by the allocation of workload $w$ (line \ref{alg:mfi:init}). For each GPU $m$ that can accommodate the requested profile $p$, we compute the current fragmentation score $F(m)$ (line \ref{alg:mfi:frag_score}).  Then, we perform dry-run allocations for each feasible profile index $i$ of profile $p$ and we store the variation of fragmentation score by computing the difference between the hypothetical fragmentation score,  ${F}^{(i)}(m)$, that would be obtained assigning profile $p$ at slice index $i$ and the original fragmentation score $F(m)$ (lines \ref{alg:mfi:dry_run_begin}-\ref{alg:mfi:dry_run_end}). 
Finally, if there are candidate GPUs, we schedule workload $w$ by allocating profile $p$ to GPU $m$ at index $i$ such that the variation in fragmentation score is minimum (\ref{alg:mfi:final_alloc_begin}-\ref{alg:mfi:final_alloc_end}).

\begin{algorithm}[t]
\caption{Minimum Fragmentation Increment (MFI)}\label{alg:mfi}
\begin{algorithmic}[1]
    \State \textbf{Input:} GPU cluster $M$, workload $w$ with MIG profile $p$
    \State \textbf{Output:} Allocation $\langle m,i \rangle, m \in M, i \in I_p$ of workload $w$ 
    \State Initialize the list $\Delta{F}^{(i)}(m) = \{\}, \forall m \in M$ \label{alg:mfi:init} 
    \For{\textbf{each} GPU $m \in M$} \label{alg:mfi:loop_gpu_begin} 
        \If{$r_w(p) \leq \Delta S_m$}  \label{alg:mfi:check_res}      
            \State Compute frag. score ${F}(m)$   \label{alg:mfi:frag_score}
            \For{\textbf{each} $i \in I_p$} \label{alg:mfi:dry_run_begin}
                \State Dry-run allocation of $p$ on index $i$ of GPU $m$ 
                \State Compute the hypothetical frag. score ${F}^{(i)}(m)$ 
                \State Store  $\Delta{{F}^{(i)}(m)}$$ \leftarrow  {F}^{(i)}(m) - {F}(m) $ 
            \EndFor \label{alg:mfi:dry_run_end} 
        \EndIf 
    \EndFor \label{alg:mfi:loop_gpu_end}
    \If{$|\Delta{F}^{(i)}(m)| \neq \{\phi\}$ } \label{alg:mfi:final_alloc_begin}
        \State Select GPU $m^*$ and index $i^*$ s.t. $\argmin \{\Delta{F}^{(i)}(m)\}$    
        \State Schedule $w$ by allocating profile $p$ to $\langle m^*,i^*\rangle$
    \Else 
        \State Reject workload $w$
    \EndIf  \label{alg:mfi:final_alloc_end}
\end{algorithmic}
\end{algorithm}

The  computational complexity required to compute the scheduling decision for a workload mainly scales with number of GPUs in the cluster. In detail, the algorithm iterates throughout all GPUs in linear time $O(kM)$, where $k$  is a constant accounting for the time required to compute fragmentation scores. Each computation depends on the number of profiles $|P|$ and MIG slice indexes $|I_p|$ supported by the GPU hardware and it is not influenced by the number of GPUs. Similarly, computing the GPU that minimizes the fragmentation variation can be done in $O(M)$ if the list of $\Delta{F}^{(i)}(m)$ values is maintained in sorted order as it is updated. Consequently, the overall complexity is $O(kM) + O(M)$, which asymptotically simplifies to $O(kM)$.

\section{Performance evaluation}

\pgfplotsset{every axis/.append style={line width=1.3pt,tick style={line width=0.4pt}} }
\pgfplotsset{every axis/.append style={font=\small}}
\pgfplotmarksize=1.5pt

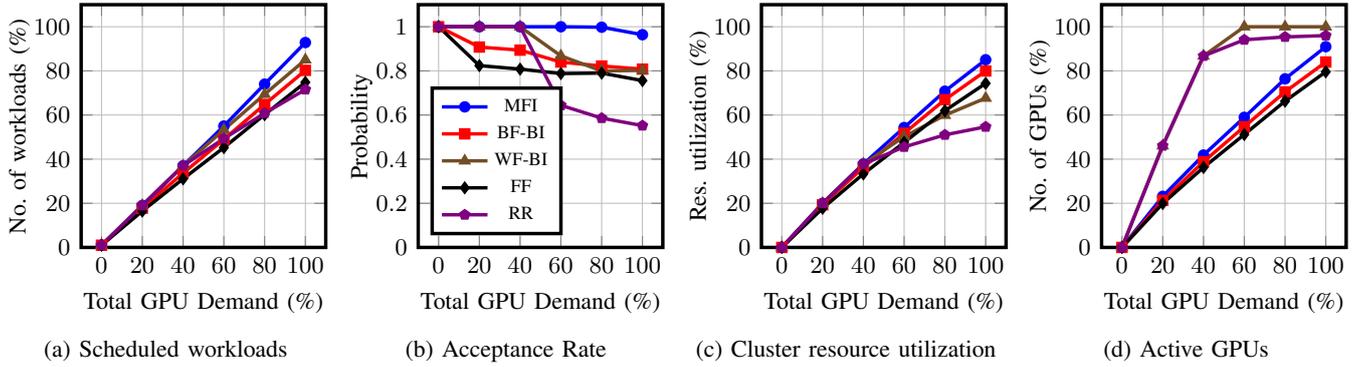
\begin{figure*}[t]
    \centering
    \begin{subfigure}[b]{0.24\textwidth}
        \centering
        \begin{tikzpicture}
            \begin{axis}[
                cycle list name=default,
                width=4.8cm, height=4.8cm,
                grid=both,
                xlabel={Total GPU Demand (\%)}, 
                xtick={0,20,40,60,80,100},
                ylabel={No. of workloads (\%)},
                ylabel style={yshift=-6pt},
                ytick={0,20,40,60,80,100},
                ymin=0,ymax=110
            ]
            \addplot table[x=x,y=MFI]{\scheduledwl};
            \addplot table[x=x,y=BF-BI]{\scheduledwl};
            \addplot table[x=x,y=WF-BI]{\scheduledwl};
            \addplot table[x=x,y=FF]{\scheduledwl};
            \addplot table[x=x,y=RR]{\scheduledwl};
            \end{axis}
        \end{tikzpicture}
        \caption{Scheduled workloads}   
        \label{subfig:uniform_dist_sched_wl}
    \end{subfigure}
    \begin{subfigure}[b]{0.24\textwidth}
        \centering
        \begin{tikzpicture}
            \begin{axis}[
                width=4.8cm, height=4.8cm,
                grid=both,
                xlabel={Total GPU Demand (\%)}, 
                xtick={0,20,40,60,80,100},
                ylabel={Probability},
                ylabel style={yshift=-6pt},
                ytick={0,0.2,0.4,0.6,0.8,1.0},
                legend style={font=\scriptsize, at={(0.05,0.05)}, anchor=south west},
                ymin=0, ymax=1.1
            ]
            \addplot table[x=x,y=MFI]{\acceptancerate};\addlegendentry{MFI}
            \addplot table[x=x,y=BF-BI]{\acceptancerate};\addlegendentry{BF-BI}
            \addplot table[x=x,y=WF-BI]{\acceptancerate};\addlegendentry{WF-BI}
            \addplot table[x=x,y=FF]{\acceptancerate};\addlegendentry{FF}
            \addplot table[x=x,y=RR]{\acceptancerate};\addlegendentry{RR}
            \end{axis}
        \end{tikzpicture}
        \caption{Acceptance Rate}
        \label{subfig:uniform_dist_prob}
    \end{subfigure}
    \begin{subfigure}[b]{0.24\textwidth}
        \centering
        \begin{tikzpicture}
            \begin{axis}[
                width=4.8cm, height=4.8cm,
                grid=both,
                xlabel={Total GPU Demand (\%)}, 
                xtick={0,20,40,60,80,100},
                ylabel={Res. utilization (\%)},
                ytick={0,20,40,60,80,100},
                ylabel style={yshift=-6pt},
                ymin=0,ymax=110
            ]
            \addplot table[x=x,y=MFI]{\resutil};
            \addplot table[x=x,y=BF-BI]{\resutil};
            \addplot table[x=x,y=WF-BI]{\resutil};
            \addplot table[x=x,y=FF]{\resutil};
            \addplot table[x=x,y=RR]{\resutil};
            \end{axis}
        \end{tikzpicture}
        \caption{Cluster resource utilization}
        \label{subfig:uniform_dist_res_util}
    \end{subfigure}
    \begin{subfigure}[b]{0.24\textwidth}
        \centering
        \begin{tikzpicture}
            \begin{axis}[
               width=4.8cm, height=4.8cm,
                grid=both,
                xlabel={Total GPU Demand (\%)}, 
                xtick={0,20,40,60,80,100},
                ylabel={No. of GPUs (\%)},
                ytick={0,20,40,60,80,100},
                ylabel style={yshift=-6pt},
                ymin=0,ymax=110
            ]
            \addplot table[x=x,y=MFI]{\activegpus}; 
            \addplot table[x=x,y=BF-BI]{\activegpus};
            \addplot table[x=x,y=WF-BI]{\activegpus}; 
            \addplot table[x=x,y=FF]{\activegpus}; 
            \addplot table[x=x,y=RR]{\activegpus};
            \end{axis}
        \end{tikzpicture}
        \caption{Active GPUs}
        \label{subfig:uniform_dist_gpus}
    \end{subfigure}
    \caption{Scheduling performance in different load conditions when MIG profile requests are uniformly distributed.}
    \label{fig:uniform_dist}
\end{figure*}


\begin{table}[t] 
\centering
\caption{MIG profile distributions}
\begin{tabular}{| c | c | c | c | c | c | c | c |}
\hline                                              
\textbf{MIG profile} & \textbf{uniform} & \textbf{skew-small} & \textbf{skew-big} &  \textbf{bimodal} \\                  
\hline
7g.80gb & 1/6 & 0.05 & 0.30 & 0.30 \\
4g.40gb & 1/6 & 0.10 & 0.25 & 0.15 \\
3g.40gb & 1/6 & 0.10 & 0.20 & 0.05 \\
2g.20gb & 1/6 & 0.20 & 0.10 & 0.05 \\
1g.20gb & 1/6 & 0.25 & 0.10 & 0.15 \\
1g.10gb & 1/6 & 0.30 & 0.05 & 0.30 \\
\hline
\end{tabular}
\label{tab2:mig_distribution}
\end{table}

We assess the performance of MFI algorithm throughout Monte Carlo simulations. In detail, we simulate the scheduling of workloads on an empty cluster of $M=100$ A100 GPUs that is progressively loaded up to its maximum capacity. To synchronize the workload duration with the scheduling procedure, we express the workload duration in terms of scheduling slots by randomly sampling a value between $[1, T]$, where $T$ correspond to the number of scheduling slots required to saturate the cluster capacity (in other words, when the total amount of requested GPU resources corresponds to the amount of resources provided by the cluster). This approach allows us to simulate the dynamics of a multi-tenant GPU cluster under varying load conditions, with workloads that have heterogeneous lifespans and arrival times.  Moreover, due to the lack of real world dataset collecting scheduling statistics of MIG-based deployments, we  analyze the performance of MFI with respect to different synthetically-generated distributions of MIG profiles to investigate their impact on the scheduling outcome. Specifically, we consider the following profile distributions, whose probability density functions are reported in Table \ref{tab2:mig_distribution}:
\begin{itemize}
    \item \textit{Uniform}: every profile is equally likely. This distribution serves as baseline for testing the scheduling performance in a general scenario.
    \item \textit{Skew-small}: small profiles are more frequent than large ones. This distribution is likely to cause severe GPU fragmentation due to the allocation of many small profiles reserving non-contiguous MIG slices which possibly deny the allocation of large profiles.
    \item \textit{Skew-big}: large profiles are more frequent than small ones. This distribution constrains scheduling flexibility, as large profiles can be placed only on a limited number of feasible MIG slice configurations. Consequently, the benefit of a fragmentation-aware scheduling policy are modest, given the inherent rigidity of the available scheduling options.
    \item \textit{Bimodal}: mixture of large and small profiles. This  distribution  highlights the scheduling performance when profiles with different amount of scheduling constraints and fragmentation generation potential are dominant. 
\end{itemize}

We compare MFI algorithm with respect to various scheduling schemes using different strategies to maximize the number of scheduled MIG profiles. In particular, to better highlight the impact of MIG scheduling over traditional scheduling techniques, we divide the benchmarks in MIG-agnostic schemes, that schedule MIG profiles based only on resource availability on the GPUs, and MIG-aware schemes, that employ a predetermined preference policy to select MIG slices indexes. In particular, we use the solution proposed by the authors of \cite{turkkan2024optimal} and we  prioritize the allocation of MIG profiles on indexes that do not restrict the placement of profiles with fewer scheduling options. For instance, the 1g.10gb profile is assigned to index 6 instead of index 0 whenever possible, thereby reserving index 0 for the 4g.40gb profile, which can only be scheduled there. Based on these considerations, we define the following benchmark schemes:
\begin{itemize}
    \item \textbf{First Fit (FF)}: this scheme is MIG-agnostic and schedule MIG profiles on the first GPU with enough available resources. Profiles are assigned to the first available index.
    \item \textbf{Round Robin (RR)}: this scheme is MIG-agnostic and sequentially distributes MIG profiles on the available GPUs. Profiles are assigned to the first available index.
    \item \textbf{Best Fit Best Index (BF-BI)}: this scheme is MIG-aware and follows a bin-packing approach which schedules MIG profiles on the GPUs minimizing the amount of available resources after the allocation. 
    \item \textbf{Worst Fit Best Index (WF-BI)}: this scheme is MIG-aware and follows a load-balancing approach which schedules MIG profiles on the GPUs maximizing the amount of available resources after the allocation. 
\end{itemize}

We quantify the performance of MFI with respect to the proposed baselines according to the following metrics: 
\begin{itemize}
    \item \textbf{Acceptance rate}: this is the probability to successfully schedule a workload on the available GPUs according to the MIG constraints. We assume that unscheduled workloads are rejected and are no longer rescheduled during the simulation. Although we acknowledge this is not realistic, this approach allows to better analyzing the performance gap between the various schemes.
    \item \textbf{Allocated workloads}: this represents the total number of successfully scheduled workloads in the GPU cluster and, together with the acceptance rate, characterizes the primary goal of the scheduling objective.
    \item \textbf{Active GPUs}: this represents the total number of GPUs required to accommodate the scheduled workloads. A GPU is considered active if it is hosting at least one workload.
    \item \textbf{Resource utilization}: this is the total amount of allocated slices in each GPU required to accommodate the various MIG profiles. Together with the number of active GPUs, this metric allows to assess the resource utilization efficiency of the various schemes.
    \item \textbf{Fragmentation severity}: this is the average fragmentation score of the cluster computed as $\frac{1}{M}\sum_{m \in M} F(m)$.

\end{itemize}
We performed 500 independent scheduling simulation for each distribution and averaging the obtained results.  All metrics are normalized with respect to  their maximum value in order to facilitate the comparison between different schemes and MIG profile distributions. 


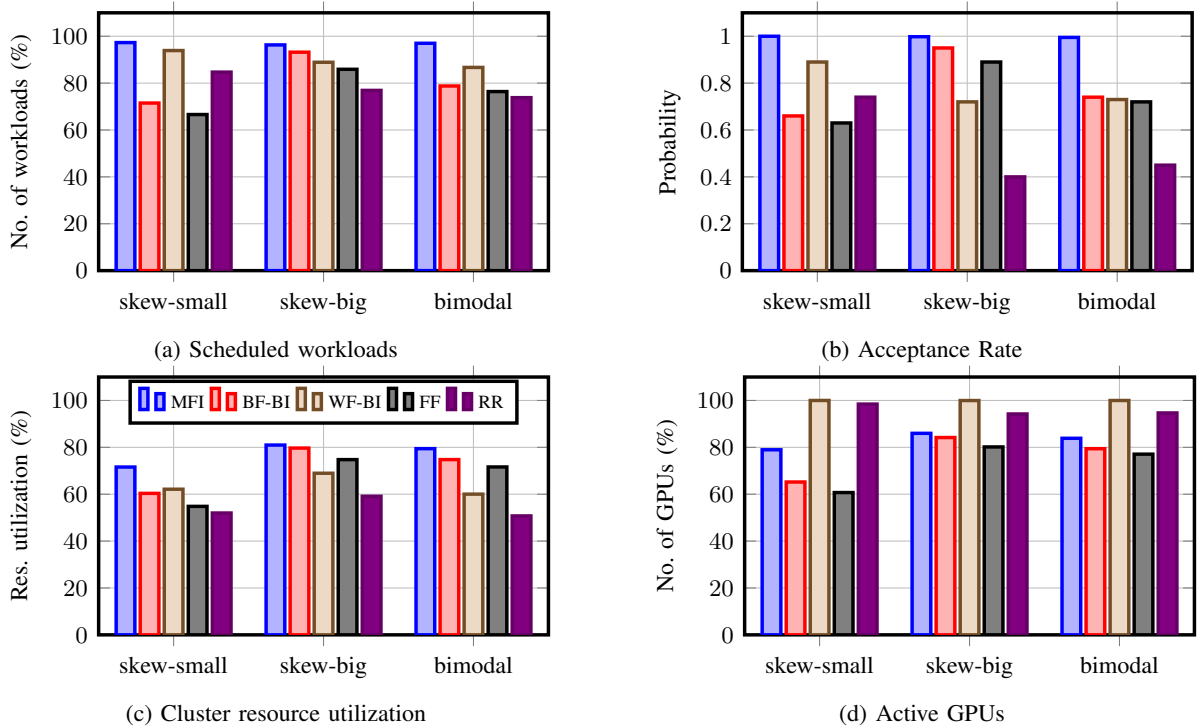
\begin{figure*}[t]
    \centering

    \begin{subfigure}[b]{0.4\textwidth}
        \centering
        \begin{tikzpicture}
        \begin{axis}[
            ybar,
            bar width=7pt,
            width=7.5cm,
            height=5cm,
            ylabel={No. of workloads (\%)},
            ytick={0,20,40,60,80,100},
            symbolic x coords={skew-small, skew-big, bimodal},
            legend style={font=\scriptsize, at={(0.5,0.99)}, anchor=north, legend columns = 5},
            xtick=data,
            grid=both,
            ymin=0,
            ymax=110,
            enlarge x limits=0.25
        ]
        \addplot coordinates {(skew-small,97.3) (skew-big,96.3) (bimodal,97.0)};
        \addplot coordinates {(skew-small,71.5) (skew-big,93.20) (bimodal,78.78)};
        \addplot coordinates {(skew-small,93.9) (skew-big,88.9) (bimodal,86.7)};
        \addplot coordinates {(skew-small,66.6) (skew-big,85.9) (bimodal,76.4)};
        \addplot coordinates {(skew-small,84.7) (skew-big,76.9) (bimodal,73.8)};
        \end{axis}
        \end{tikzpicture}
        \caption{Scheduled workloads}
        \label{subfig:all_dist_heavy_sched_wl}
    \end{subfigure}
    \hspace{1cm}
    \begin{subfigure}[b]{0.4\textwidth}
        \centering
        \begin{tikzpicture}
        \begin{axis}[
            ybar,
            bar width=7pt,
            width=7.5cm,
            height=5cm,
            ylabel={Probability},
            ytick={0,0.20,0.40,0.60,0.80,1},
            symbolic x coords={skew-small, skew-big, bimodal},
            xtick=data,
            grid=both,
            ymin=0,
            ymax=1.1,
            enlarge x limits=0.25
        ]
        \addplot coordinates {(skew-small,1) (skew-big,0.998) (bimodal,0.995)};
        \addplot coordinates {(skew-small,0.66) (skew-big,0.95) (bimodal,0.74)};
        \addplot coordinates {(skew-small,0.89) (skew-big,0.72) (bimodal,0.73)};
        \addplot coordinates {(skew-small,0.63) (skew-big,0.89) (bimodal,0.72)};
        \addplot coordinates {(skew-small,0.74) (skew-big,0.4) (bimodal,0.45)};
        \end{axis}
        \end{tikzpicture}
        \caption{Acceptance Rate}
        \label{subfig:all_dist_heavy_acceptance_rate}
    \end{subfigure}
    \begin{subfigure}[b]{0.4\textwidth}
        \centering
        \begin{tikzpicture}
        \begin{axis}[
            ybar,
            bar width=7pt,
            width=7.5cm,
            height=5cm,
            ylabel={Res. utilization (\%)},
            symbolic x coords={skew-small, skew-big, bimodal},
            ytick={0,20,40,60,80,100},
            xtick=data,
            grid=both,
            legend style={font=\scriptsize, at={(0.5,0.99)}, anchor=north, legend columns = 5},
            ymin=0,
            ymax=110,
            enlarge x limits=0.25
        ]
        \addplot coordinates {(skew-small,71.5908) (skew-big,80.96966) (bimodal,79.43312)};
        \addplot coordinates {(skew-small,60.3637) (skew-big,79.68178) (bimodal,74.7701)};
        \addplot coordinates {(skew-small,62.13132) (skew-big,68.94866) (bimodal,60.03112)};
        \addplot coordinates {(skew-small,54.81354) (skew-big,74.763) (bimodal,71.62648)};
        \addplot coordinates {(skew-small,51.98286) (skew-big,59.11872) (bimodal,50.73672)};
        \legend{MFI,BF-BI,WF-BI,FF,RR}
        \end{axis}
        \end{tikzpicture}
        \caption{Cluster resource utilization}
        \label{subfig:all_dist_heavy_res_util}
    \end{subfigure}
    \hspace{1cm}
    \begin{subfigure}[b]{0.4\textwidth}
        \centering
        \begin{tikzpicture}
        \begin{axis}[
            ybar,
            bar width=7pt,
            width=7.5cm,
            height=5cm,
            ylabel={No. of GPUs (\%)},
            symbolic x coords={skew-small, skew-big, bimodal},
            ytick={0,20,40,60,80,100},
            xtick=data,
            grid=both,
            ymin=0,
            ymax=110,
            enlarge x limits=0.25
        ]
        \addplot coordinates {(skew-small,78.96) (skew-big,85.975) (bimodal,83.86)};
        \addplot coordinates {(skew-small,65.205) (skew-big,84.185) (bimodal,79.41)};
        \addplot coordinates {(skew-small,100) (skew-big,99.97) (bimodal,99.99)};
        \addplot coordinates {(skew-small,60.71) (skew-big,80.17) (bimodal,77.09)};
        \addplot coordinates {(skew-small,98.45) (skew-big,94.235) (bimodal,94.675)};
        \end{axis}
        \end{tikzpicture}
        \caption{Active GPUs}
        \label{subfig:all_dist_heavy_active_gpus}
    \end{subfigure}
    
    \caption{Scheduling performance for different distributions of MIG profiles when the amount of requested GPU resources is 85\% of the cluster capacity.}
    \label{fig:all_dist_heavy}
\end{figure*}


We initially discuss the results by  analyzing the performance of the various schemes when MIG profile requests are uniformly distributed. In Fig. \ref{fig:uniform_dist}, we report the considered metrics as the number of requested resources to the GPU cluster increases. For example, a GPU demand of 50\% corresponds to the scenario where the total amount of workloads arrived between slot 0 and slot $t$ requires 50\% of GPU resources of the cluster in order to be accommodated. Note that this value does not account for workloads termination freeing up resources, hence the actual resource utilization of the cluster is lower. 
We observe that MFI achieves the highest number of scheduled workloads (Fig. \ref{subfig:uniform_dist_sched_wl}) by maintaining an acceptance rate close to 100\% regardless of the cluster load (Fig. \ref{subfig:uniform_dist_prob}). This is because it adapts the allocation of MIG profiles to counteract the increase of fragmentation caused by the combination of rigid MIG constraints and elastic resource availability. Conversely, the various baselines have visibly worse performance. In particular, the MIG-agnostic schemes FF and RR perform worst due to the allocation of MIG profiles on unfeasible indexes. This effect is less severe for low resource loads since more scheduling options are available, hence it favors RR scheme which spreads profiles on unused GPUs. However its performance sharply deteriorates as the cluster utilization increases, exacerbating the effect of fragmentation and making the successful allocation of workloads more challenging. In this condition, FF performs better due its higher bin-packing capability, ensuring a higher acceptance rate in heavy load conditions. MIG-aware schemes behave similar to the MIG-agnostic counterparts, but provide better results, confirming that the preference-based index policy  is an effective mechanism to handle MIG constraints. However, although this strategy aids the fragmentation mitigation, the static nature of the policy makes the scheduling strategy unable to adapt to the different fragmentation scenarios caused by the continuous workload arrival/termination. In regard of resource efficiency, we observe that MFI achieves the highest resource utilization (Fig. \ref{subfig:uniform_dist_res_util}) employing a proportionate number of GPUs (Fig. \ref{subfig:uniform_dist_gpus}) as the cluster load increases. In contrast, RR and WF-BI quickly saturate the available GPUs due to their spreading strategy, which helps satisfying MIG constraints but also requires a higher number of  GPUs. Conversely, FF and BF-BI pack MIG profile requests more efficiently, with BF-BI achieving greater improvements thanks to its MIG-aware scheduling. However, the smaller number of GPUs they employ is not utilized effectively because higher fragmentation reduces overall resource usage. This stands in contrast to MFI, which achieves higher utilization by limiting fragmentation.

We extend the results analysis by showing the performance obtained with different distributions of MIG profile requests. In Fig. \ref{fig:all_dist_heavy}, we provide a snapshot of the performance on a heavy load scenario corresponding to resource request that is 85\% of the cluster capacity. We observe that MFI outperforms the considered baselines regardless of the MIG profile distribution  and provides consistent results. In detail, on average, the performance gap is wider when MIG profiles follows the \textit{skew-small} distribution. The high number of small profiles creates frequent non-contiguous allocation of MIG profiles, thus leading to severe fragmentation. Bin packing strategies like BF-BI and FF are most affected as they prioritize the allocation of most utilized GPUs which are also likely to be the most fragmented. In this context, the fixed index selection policy used by BF-BI provides limited benefit since big profiles are rarely scheduled. As results, the number of scheduled workloads (Fig. \ref{subfig:all_dist_heavy_sched_wl}) and acceptance rate (Fig. \ref{subfig:all_dist_heavy_acceptance_rate}) are considerably lower than other baselines strategies which instead distribute workloads across GPUs, delaying the fragmentation growth. In contrast, the performance gap with respect to MFI is smaller when profiles are distributed according to \textit{skew-big}. The prevalence of big MIG profile requests reduces the fragmentation chances due to the limited number of scheduling options that essentially force the workload allocation (for example, 7.80gb requires a full GPU and 4g.40gb can only be allocated on index 0). Therefore, contrary to the previous case, FF and BF-BI achieve a better resource utilization (Fig. \ref{subfig:all_dist_heavy_res_util}) compared RR and WF-BI schemes using a lower amount of GPUs (Fig. \ref{subfig:all_dist_heavy_active_gpus}). Finally, when MIG profiles follows a \textit{bimodal} distribution, the combination of small and big profiles increases the scheduling complexity due to the presence of conflicting MIG constraints, which hinder the workload acceptance rate performance of all baselines. Nonetheless, MFI provides steady performance thanks to its proactive scheduling approach which suppresses the fragmentation growth, achieving higher acceptance rates with a limited number of GPUs. 

\begin{figure}[!t]
    \centering
    \definecolor{color1}{RGB}{31,119,180} 
    \definecolor{color2}{RGB}{255,127,14} 
    \definecolor{color3}{RGB}{44,160,44}  
    \definecolor{color4}{RGB}{214,39,40}  
    \definecolor{color5}{RGB}{148,103,189} 
    \begin{tikzpicture}
    \begin{axis}[
        ybar,
        bar width=7pt,
        width=9cm,
        height=5cm,
        grid=both,
        ylabel={Fragmentation score},
        ylabel style={yshift=-6pt},
        symbolic x coords={uniform, skew-small, skew-big, bimodal},
        legend style={font=\scriptsize, at={(0.55,0.98)}, anchor=north, legend columns = 5},
        xtick=data,
        ymin=0,ymax= 9,
        enlarge x limits=0.15,
    ]
    \addplot coordinates {(uniform,0.4142780618) (skew-small,1.018375293) (skew-big,0.1528759404) (bimodal,0.4571790065)};
    \addplot coordinates {(uniform,0.6867703212) (skew-small,1.469294706) (skew-big,0.2925574473) (bimodal,0.5770162889)};
    \addplot coordinates {(uniform,1.622476706) (skew-small,2.986084681) (skew-big,0.7424384059) (bimodal,1.665531417)};
    \addplot coordinates {(uniform,1.314443037) (skew-small,2.411984874) (skew-big,0.7731936581) (bimodal,1.099762067)};
    \addplot coordinates {(uniform,5.353519804) (skew-small,7.114125416) (skew-big,3.476620554) (bimodal,4.917497222)};
    \legend{MFI,BF-BI,WF-BI,FF,RR}
    \end{axis}
    \end{tikzpicture}
    \caption{Average fragmentation score obtained by different scheduling strategies and MIG profiles distributions.}
    \label{fig:frag_score}
\end{figure}
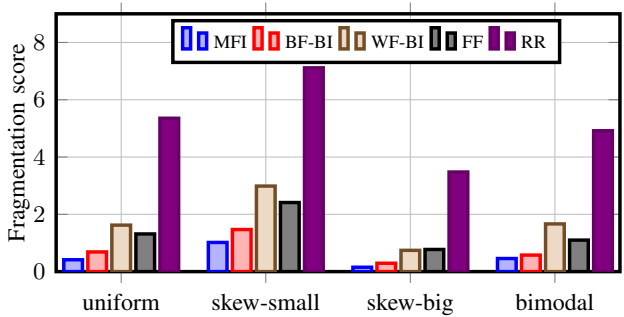

This analysis demonstrates that the proposed fragmentation score metric can be used to consistently drive the workload scheduling process towards a resource-efficient utilization of GPU resources regardless of the MIG profiles distribution. To further  validate this argument, in Fig. \ref{fig:frag_score}, we show the average fragmentation score obtained by the various schemes for each considered profile distribution. As expected, the improved performance achieved with MFI corresponds to lower fragmentation score values. Similarly, the fragmentation scores for the other baselines are consistent with their respective performance: lower acceptance rates and lower resource utilization correspond to higher fragmentation scores, and vice versa.

\section{Conclusion}

We addressed the problem of GPU resource fragmentation arising from the rigid constraints imposed by MIG and the continuous allocation and release of GPU resources over time. This phenomenon significantly reduces the scheduling efficiency of multi-tenant GPU clusters, where workload isolation is essential to guarantee both quality of service and security of co-located tenants. In this context, we proposed a fragmentation metric tailored for MIG to quantify the fragmentation severity of GPUs and we leveraged this information to drive the scheduling decision of incoming workloads. In detail, the designed fragmentation-aware scheduling algorithm operates in an online manner, without any a-priori knowledge of the workload distribution. It adopts a proactive allocation strategy to suppress the fragmentation growth on GPUs, making possible to increase the number of successfully allocated workloads through more efficient resource utilization.  We evaluated the proposed solution against multiple baselines implementing both MIG-agnostic and MIG-aware scheduling policies, as well as across various MIG profile distributions to assess performance sensitivity. Results reveal that our scheme consistently achieves the highest workload acceptance rate under different load conditions while ensuring an efficient use of GPU resources.

\section*{Acknowledgment}
This work was supported by Ministero delle Imprese e del Made in Italy (IPCEI Cloud DM 27 giugno 2022 - IPCEI-CL-0000007) and European Union (Next Generation EU).

\bibliographystyle{IEEEtran}
\bibliography{bibliography}

\end{document}